\documentclass[aps,reprint,prb]{revtex4-2}
\usepackage{amsmath}
\usepackage{amssymb}
\usepackage{graphicx}
\begin{document}
\title{Multiferroic nitride perovskites with giant polarizations and large magnetic moments}
\author{Churen Gui}
\author{Jun Chen}
\author{Shuai Dong}
\email{Corresponding author. Email: sdong@seu.edu.cn}
\affiliation{School of Physics, Southeast University, Nanjing 211189, China}
\begin{abstract}
Multiferroics with coupling between ferroelectricity and magnetism have been pursued for decades. However, their magnetoelectric performances remain limited due to the common trade-off between ferroelectricity and magnetism. Here, a family of nitride perovskites is proposed as multiferroics with prominent physical properties and nontrivial mechanisms. Taking GdWN$_3$ as a prototype, our first-principles calculations found that its perovskite phases own large polarizations (e.g. $111.3$ $\mu$C/cm$^2$ for the $R3c$ phase) and a magnetic moment $7$ $\mu_{\rm B}$/Gd$^{3+}$. More interestingly, its ferroelectric origin is multiple, with significant contributions from both Gd$^{3+}$ and W$^{6+}$ ions, different from its sister member LaWN$_3$ in which the ferroelectricity almost arises from W$^{6+}$ ions only. With decreasing size of rare earth ions, the A site ions would contribute more and more to the ferroelectric instability. Considering that small rare earth ions can be primary origins of both proper ferroelectricity and magnetism in nitride perovskites, our work provides a route to pursuit more multiferroics with unconventional mechanisms and optimal performances.
\end{abstract}
\maketitle

\section{Introduction}
Multiferroics mostly refer to those materials simultaneously exhibiting ferroelectric order and magnetic order, which have been extensively studied for decades \cite{Dong2015}. The essential issue in this topic is to pursuit the mutual manipulation of these two orders \cite{Tokura2014,Dong2019,Spaldin2021}. For applications, the desired properties for an ideal multiferroic material include a large polarization, a large magnetization, and strong coupling between them. However, the current available multiferroics, mostly based on oxides, can not satisfy these conditions simultaneously. Instead, in most cases they are mutually exclusive. For example, in the typical type-I multiferroic BiFeO$_3$, its ferroelectricity and antiferromagnetism are both prominent luckily, but their difference sources (Bi$^{3+}$ {\it vs} Fe$^{3+}$) make their coupling naturally indirect and weak \cite{Wang2003,Zhang2011,Burns2019}. In contrast, in those so-called type-II multiferroics like TbMnO$_3$ \cite{Kimura2003,Dong2012,Schoenherr2020}, the magnetoelectric couplings can be intrinsically strong, which are valuable regarding the magnetic control of ferroelectricity. However, their improper ferroelectric polarizations are typically weak.

Thus, it is interesting to search for multiferroics with same-ion-rooted large polarization and strong magnetism, which may provide alternative mechanisms to solve aforementioned dilemma. Sr$_{1-x}$Ba$_x$MnO$_3$ is an example in this category \cite{Rondinelli2009,Sakai2011}. Its polarization can reach $25$ $\mu$C/cm$^2$ driven by the 2nd order Jahn-Teller distortion of Mn$^{4+}$, which also contributes to the G-type antiferromagnetism with $3$ $\mu_{\rm B}$/Mn. However, only a few oxides exhibit the same-ion-rooted multiferroicity, and those compounds with similar mechanism but larger polarization and magnetic moments are certainly more attractive.

Recent advances in nitride perovskites have attracted many attention and revealed that, comparing to oxide analogues, nitrides could exhibit more excellent ferroelectric properties \cite{Sarmiento-Perez2015,Korbel2016,Fang2017,Gui2020,Talley2021}. A natural advantage is that the high negative/positive valences of nitrogen/metal ions can result in giant polarizations. Very recently, the successful synthesis of polar nitride perovskite LaWN$_3$ inspired the community \cite{Talley2021}, and encourage further studies to find more candidates, both theoretically and experimentally. Although the high valence of B site transition metal ion may exclude the magnetic moment, its A site ion can provide the possibility to obtain magnetism in nitride perovskites. Indeed, a previous work on $R$ReN$_3$ and $R$WN$_3$ ($R$: rare earth) revealed the rare earth magnetism \cite{Flores-Livas2019}. However, most $R$ReN$_3$ are non-polar perovskites and most $R$WN$_3$ even have non-perovskite ground states. So the polar perovskites $RB$N$_3$ still need further studies, and the multiferroicity of $RB$N$_3$ has not been discovered yet.

In this work, we will investigate an example of nitride perovskite, GdWN$_3$, to elucidate the multiferroicity of $RB$N$_3$.

\begin{figure}
\centering
\includegraphics[width=0.48\textwidth]{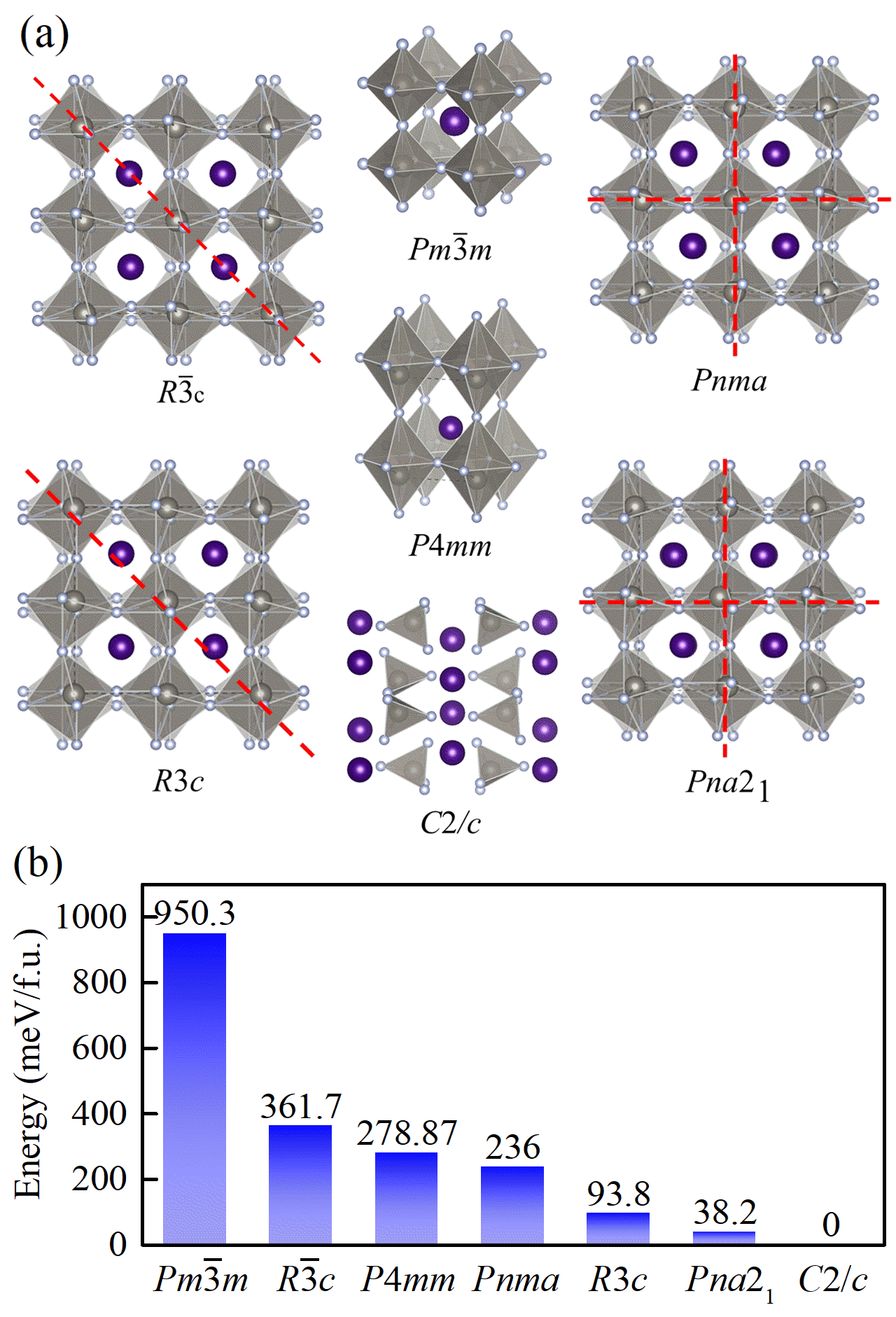}
\caption{(a) Various crystal structures of GdWN$_3$. The distorted perovskite phases are shown in their pseudocubic cells, and the red lines indicate the mirror plane of centrosymmetric structures ($Pnma$ and $R\bar3c$) to emphasize the ferroelectric distortions in corresponding ferroelectric $Pna2_1$ and $R3c$ phases. (b) Their relative energies with the $C2/c$ phase as the reference (i.e. $0$). The structures and energies of the $C2/m$ and $P2_1/c$ phases are not shown, since their energies are too high ($1.5$ eV/f.u.).}
\label{fig1}
\end{figure}

\section{Methods}.
Our density functional theory (DFT) calculations are performed using Vienna {\it ab initio} Simulation Package (VASP) \cite{Kresse1996}. The plane-wave cutoff energy is $500$ eV, and $7\times7\times7$ Monkhorst-Pack $\Gamma$-centered $k$-point mesh is used. The convergence criteria for electronic iteration and structural relaxations is set to be $10^{-5}$ eV and $10^{-3}$ eV/{\AA}, respectively. The ferroelectric polarization is calculated using the Berry phase method \cite{King-Smith1993,Resta1994}. The strongly constrained and appropriately normed (SCAN) density functional is used as it's supposed to be superior to most gradient corrected functionals \cite{Sun2015,Sun2016}, which can lead to similar results to the conventional GGA+$U$ correction to Gd's $4f$ orbitals (details of comparison can be found in Supplemental Materials (SM) \cite{supp}).

The structural dynamic stability is verified by the vibrational spectra, obtained based on the density functional perturbation theory (DFPT) \cite{Gonze1997}. Phonopy is adopted to calculate the phonon band structures \cite{Togo2015}, and the AFLOW is used to seek and visualize the dispersion paths in Brillouin zone \cite{Curtarolo2012}. Crystal structures are visualized using VESTA \cite{Momma2011}.

Moreover, to estimate the magnetic transition temperatures, the Monto Carlo (MC) simulations based on Heisenberg model are performed. The simulations adopt a $18\times18\times18$ lattice with periodic boundary condition, and larger lattices were tested to comfirm the results. The first $3\times10^4$ MC steps (MCSs) are used for thermal equilibrium, then another $3\times10^4$ MCSs are used for measurement. Then the specific heat is used to indicate the phase transition point.

\section{Results \& Discussion}
\subsection{Structures \& ferroelectricity}
Based on previous studies on nitride perovskites \cite{Fang2017,Flores-Livas2019,Gui2020}, six possible perovskite phases are considered for GdWN$_3$: $Pm\bar3m$, $R3c$, $R\bar3c$, $Pna2_1$, $P4mm$, and $Pnma$, as well as three non-perovskite phases: $C2/c$, $C2/m$, and $P2_1/c$ \cite{Flores-Livas2019,Sarmiento-Perez2015}. The $Pm\bar3m$ phase is the ideal undistorted cubic perovskite, as a parent structure for all distorted perovskites. These crystal structures and corresponding DFT energies are shown in Fig.~\ref{fig1}.

The $C2/c$ phase owns the lowest energy, in agreement with the previous study \cite{Flores-Livas2019}. And the $Pm\bar3m$ phase owns a much higher energy than all other perovskites, indicating that severe distortions will exist in the perovskite structure. Such spontaneous distortions can be attributed to the low tolerance factor of GdWN$_3$ ($0.881$, much lower than $0.969$ for LaWN$_3$). Even though, the energy differences between the $C2/c$ phase and polar perovskite phases (orthorhombic $Pna2_1$ and rhombohedral $R3c$), are less than $100$ meV/f.u., leaving promsing possibility to stabilize these phases. The methods to stabilize the perovskite phases will be discussed later, which are helpful guides for following experiments.

\begin{figure}
\centering
\includegraphics[width=0.48\textwidth]{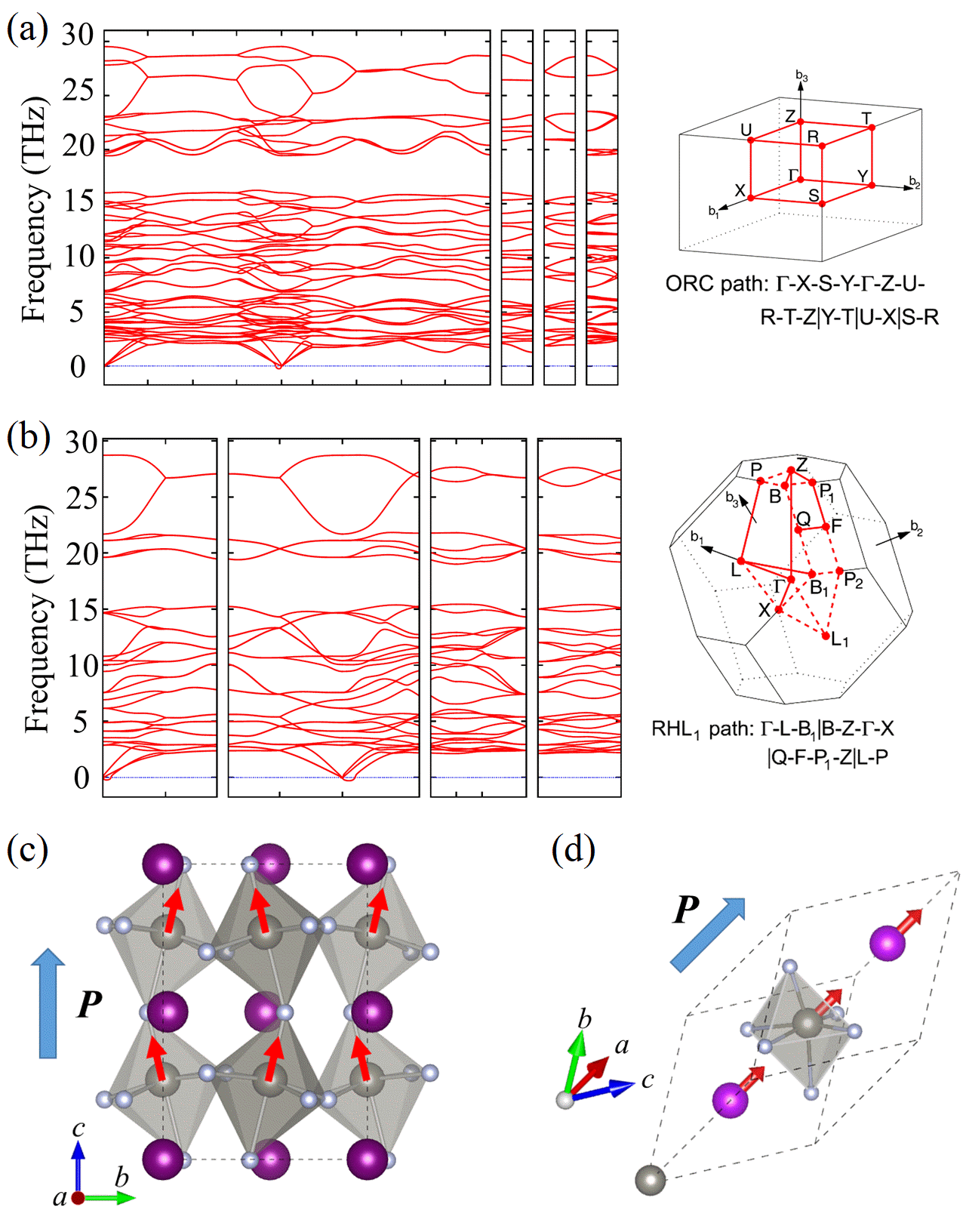}
\caption{(a-b) The phonon spectra of the $Pna2_1$ and $R3c$ phases. The corresponding Brillion zone and the dispersion paths (red lines) between high-symmetry points are also indicated. (c-d) Schematic of ferroelectric structures. Red arrows: the ion displacements of W$^{6+}$ and Gd$^{3+}$ from the centrosymmetric positions; Blue arrows indicate the directions of polarizations: along the [001] axis for $Pna2_1$ and [111] axis for $R3c$. }
\label{fig2}
\end{figure}

In real materials, metastable phases with slightly higher energies may also exist in ambient condition, e.g. diamond. The structural dynamic stability is a neccessary criterion. Thus the phonon spectra of $Pna2_1$ and $R3c$ phases are calculated, as shown in Fig.~\ref{fig2}(a-b). Imaginary vibration mode does not exist in either case, indicating their dynamic stability.

For perovskites, the distortion of octahedra are essential to determine their physical properties, especially its polarity. Using the Glazer's notation \cite{Glazer1972}, the tilting and rotation of WN$_6$ octahedra in the $Pna2_1$ phase is describled as $a^-a^-c^+$. Within the octahedral cage, W$^{6+}$ ion moves to the upper (or lower) N$^{3-}$ ion, as shown in Fig.~\ref{fig2}(c). Such displacements of W$^{6+}$ ions result in a net dipole pointing along the [001] direction of pseudocubic cell. Its ferroelectric polarization ($P$) is estimated as $52$ $\mu$C/cm$^2$, much larger than that of $Pna2_1$ LaWN$_3$ ($20$ $\mu$C/cm$^2$ \cite{Fang2017}).

For the rhombohedral $R3c$ phase, its tilting and rotation mode is $a^-a^-a^-$. The ferroelectric displacements attributed to both W$^{6+}$ and Gd$^{3+}$ ions moving to the diagonal direction of the octahedron, i.e., one of the eightfold $<111>$ directions of the pseudocubic cell, as shown in Fig.~\ref{fig2}(d). Its ferroelectric $P$ is very large ($111.3$ $\mu$C/cm$^2$), even larger than the $R3c$ BiFeO$_3$ ($\sim90$ $\mu$C/cm$^2$ \cite{Wang2003,Choi2009}) and $R3c$ LaWN$_3$ ($84$ $\mu$C/cm$^2$ in our calculation).

\begin{figure}
\centering
\includegraphics[width=0.48\textwidth]{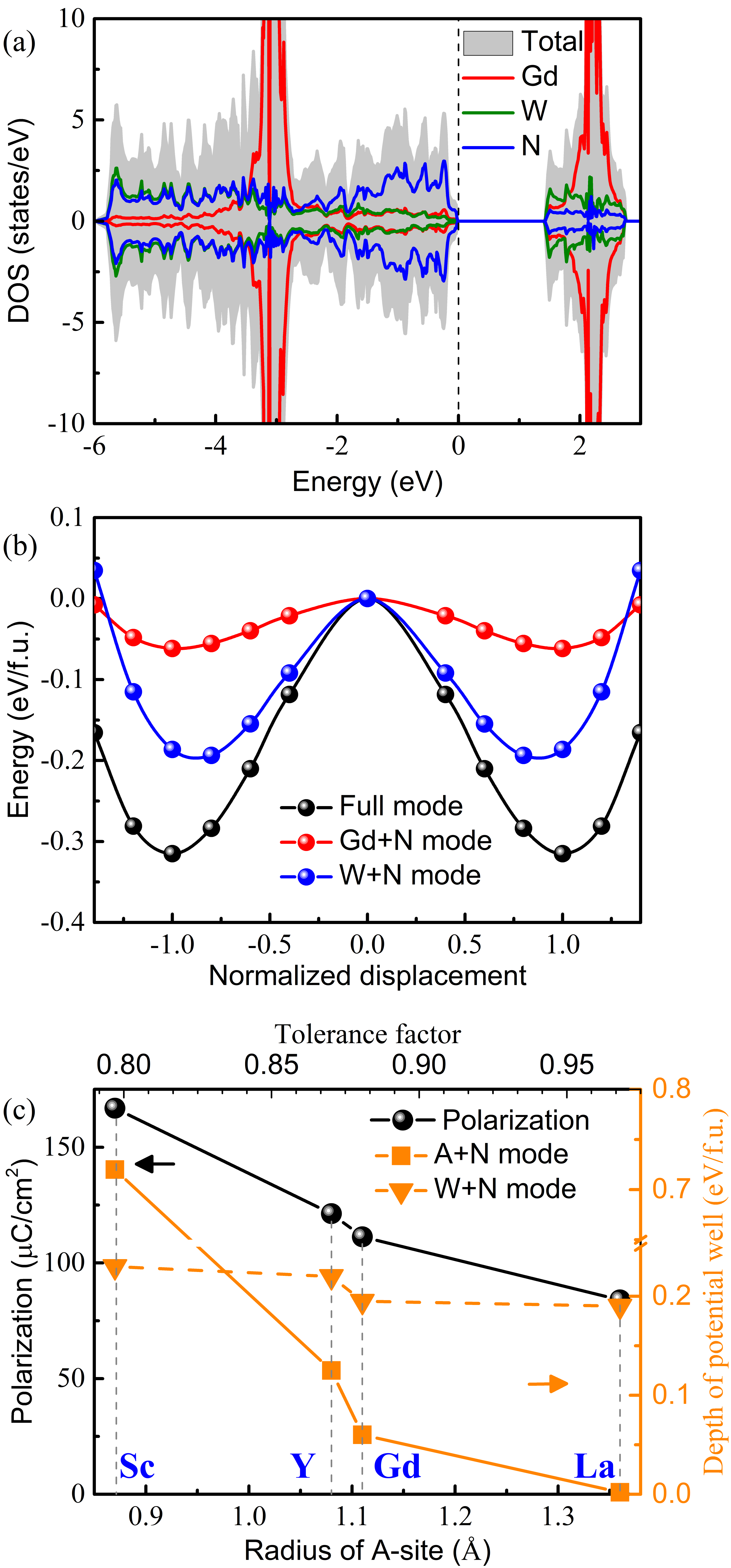}
\caption{(a) The atom-projected electronic DOS for $R3c$ GdWN$_3$. Strong hybridization between N's $2p$ and W's $5d$ orbitals can be clearly evidenced, although the $5d$ orbitals of W$^{6+}$ are nominally empty. (b) Double-well energy profiles as a function of normalized displacement for the full mode and partial modes. The optimized $R3c$ phase is set at $1$ and paraelectric $R\bar3c$ phase locates at $0$. (c) The polarization and the well-depth of energy profiles of A+N and W+N modes as a function of radius of A site ion.}
\label{fig3}
\end{figure}

The ferroelectric instability in LaWN$_3$ was claimed to originate from the strong hybridization between N's $2p$ and W's $5d$ orbitals \cite{Fang2017}. Such hybridization also presents in GdWN$_3$, as evidenced in its electronic densities of states (DOS) [Fig.~\ref{fig3}(a)]. However, this hybridization can not quantitatively explain why the value of $P$ in GdWN$_3$ is much larger than that in LaWN$_3$.

The ferroelectric switching barrier of $R3c$ GdWN$_3$ is calculated, which reaches $0.32$ eV/f.u., as shown in Fig.~\ref{fig3}(b). Such a deep double-well profile (the full mode) implies a strong tendency of ferroelectricity. Although in principle the DFT method itself can not estimate the precise ferroelectric Curie temperature ($T_{\rm C}$), a high $T_{\rm C}$ above room temperature is highly promising for $R3c$ GdWN$_3$. For reference, the ferroelectric energy well for $R3c$ BiFeO$_3$ is $0.43$ eV/f.u. and its ferroelectric $T_{\rm C}$ reaches $1103$ K. Furthermore, the large switching barrier corresponds to a coercive field of $\sim 7.7\times10^8$ V/m, which is acceptable since the value is comparable to that of BiFeO$_3$ ($\sim 1.2\times10^9$ V/m).

To claraify the origin of its giant $P$, the energy gains of partial distortion modes (Gd+N and W+N) are calculated, as compared in Fig.~\ref{fig3}(b). These partial distortion modes denote the displacements of selected ions only, while the full mode denotes the displacements of all ions. Interestingly, similar double-well energy profiles are also observed for the partial Gd+N and W+N modes, although the former is shallower than the latter. This dual double-well profiles imply that not only the W-N orbital hybridization contributes to its polarization, but also the Gd ion also has prominent contribution even it is secondary. In contrast, in $R3c$ LaWN$_3$, the depth of energy well for the La+N partial mode is almost zero: only $0.002$ eV/f.u. (in the magnitude of DFT precision), as shown in Fig. S1 in SM \cite{supp} and in agreement with Ref.~\cite{Fang2017}. In other words, La$^{3+}$ ion only plays a passive role in its ferroelectric transition, which is rather common in ferroelectric oxide perovskites. In this sense, the ferroelectric origin in GdWN$_3$ is nontrivial.

In oxide perovskites, some ions like Pb$^{2+}$ and Bi$^{3+}$ can induce ferroelectricity due to their $6s^2$ lone pairs \cite{Spaldin2021,Ghita2005}. However, there is no such $6s^2$ lone pair in Gd$^{3+}$. Then why can it be ferroelectric active here? The reason is the low tolerance factor due to the small A site ions: Gd$^{3+}$ ($1.11$ \AA) is smaller than La$^{3+}$ ($1.36$ \AA) \cite{Singh2006}. A small ion in the center of a large cavity may be dynamically unstable, and thus the spontaneous displacement from the center can strength the bonding energy. Similar situation occurs in so-called ferroelectric metal LiOsO$_3$, where Li$^+$ plays this role \cite{Liu2015}.

To futher confirm this mechanism, we replace Gd$^{3+}$ with even smaller Y$^{3+}$ ($1.08$ \AA) and Sc$^{3+}$ ($0.87$ \AA). Their optimized $R3c$ structures remain dynamically stable, as shown in Fig.~S2 of SM \cite{supp}. These smaller A site ions decrease the cell volume and stretch the rhombohedral $R3c$ cell along the $<111>$ direction, i.e. the polarization direction. Larger polar displacements occurs, resulting in even larger polarizations: $121.3$ $\mu$C/cm$^2$ for YWN$_3$ and $166.8$ $\mu$C/cm$^2$ for ScWN$_3$. Simiar analysis of their energy profiles of partial modes are shown in Fig.~S2 of SM \cite{supp}. The extracted well depths of A site+N and W+N modes are compared in Fig.~\ref{fig3}(c). It is obvious that the well depth of W+N mode is always $\sim0.2$ eV/f.u., independent on the size of A site ion. In contrast, smaller A site ions trigger the A+N mode, whose potential well increases rapidly with decreasing A site size. All these evidences support the nontrivial ferroelectricity that small Gd$^{3+}$ ions trigger the additional A site contribution, which not only strengthens the net polarization but also provides a route to strong magnetoelectricity.

\begin{table}
\centering
\setlength{\tabcolsep}{0.01\textwidth}
\caption{DFT results of two polar phases of GdWN$_3$, including the magnetic interactions (in unit of meV), band gaps $E_g$(in unit of eV), and ferroelectric polarizations $P$ (in unit of $\mu$C/cm$^2$). Due to the symmetry requirement, the effective exchange $J$ is isotropic in the $R3c$ phase, but is anisotropic in the $Pna2_1$ phase (thus the indices ($x$, $y$, $z$) indicate the direction of the exchange interaction). And its magnetic hard axis is along the polarization direction (i.e., the [111] axis of the pseudo cubic framework), which is chosen as $z$ axis here. Here the spin is normalized for simplicity (i.e., $|\mathbf{S}|=1$).}
\begin{tabular*}{0.48\textwidth}{lccccccc}
\hline \hline
Phase & $J_x$ & $J_y$ & $J_z$ & $A_x$ & $A_z$ & $E_g$ & $P$\\
\hline
$R3c$ & $0.34$ & $0.34$ & $0.34$ & $0$ & $0.08$ & $1.42$ &$111.3$\\
$Pna2_1$ & $0.45$ & $0.46$ & $0.39$ & $-0.02$ & $-0.06$ & $1.13$ &$52.0$\\
\hline \hline
\end{tabular*}
\label{table}
\end{table}

\subsection{Magnetism \& magnetoelectricity}
Our calculation confirms that the magnetic moment of Gd$^{3+}$ is a large value $\sim7$ $\mu_{\rm B}$ in all phases, as expected for its half-filled $4f$ orbitals, in consistent with the large value reported in non-perovskite GdWN$_3$ \cite{Flores-Livas2019}. According to the DFT energy comparison, the magnetic ground state is the G-type antiferromagnetism in both the $Pna2_1$ and $R3c$ phases, with all spin moments align antiparallelly with their nearest neighbors. To describe this spin lattice, the Heisenberg spin model is adopted, which can be expressed as:
\begin{equation}
\centering
H=\sum_{<ij>}J_{ij}\mathbf{S}_i\cdot\mathbf{S}_j+\sum_i[A_z(S_i^z)^2+A_x(S_i^x)^2],
\label{hami}
\end{equation}
where the first item is the effective exchange interaction. Since the $4f$ electrons are highly localized, the exchange interactions ($J$'s) between nearest-neighboring Gd's spins are naturally weak. The second item is the magnetic anisotropy, and a positive $A_z$ ($A_x$) implies the magnetic hard axis. Based on the DFT energies, the coefficients $J$'s and $A$'s are estimated (see calculation details in SM \cite{supp}), as summarized in Table~\ref{table}.

The N\'eel temperatures ($T_N$) are estimated using the MC simulation: $\sim7.4$ K for the $Pna2_1$ phase and $\sim5.8$ K for the $R3c$ phase, as shown in Fig.~S3 of SM \cite{supp}. Comparing to those $3d$ magnets, $J$'s of $4f$ magnets are much smaller. Thus $T_N$'s of GdWN$_3$ in both phases are typically low, e.g. much lower than that of BiFeO$_3$ ($\sim607$ K \cite{Xu2019a}).

\begin{figure}
\centering
\includegraphics[width=0.48\textwidth]{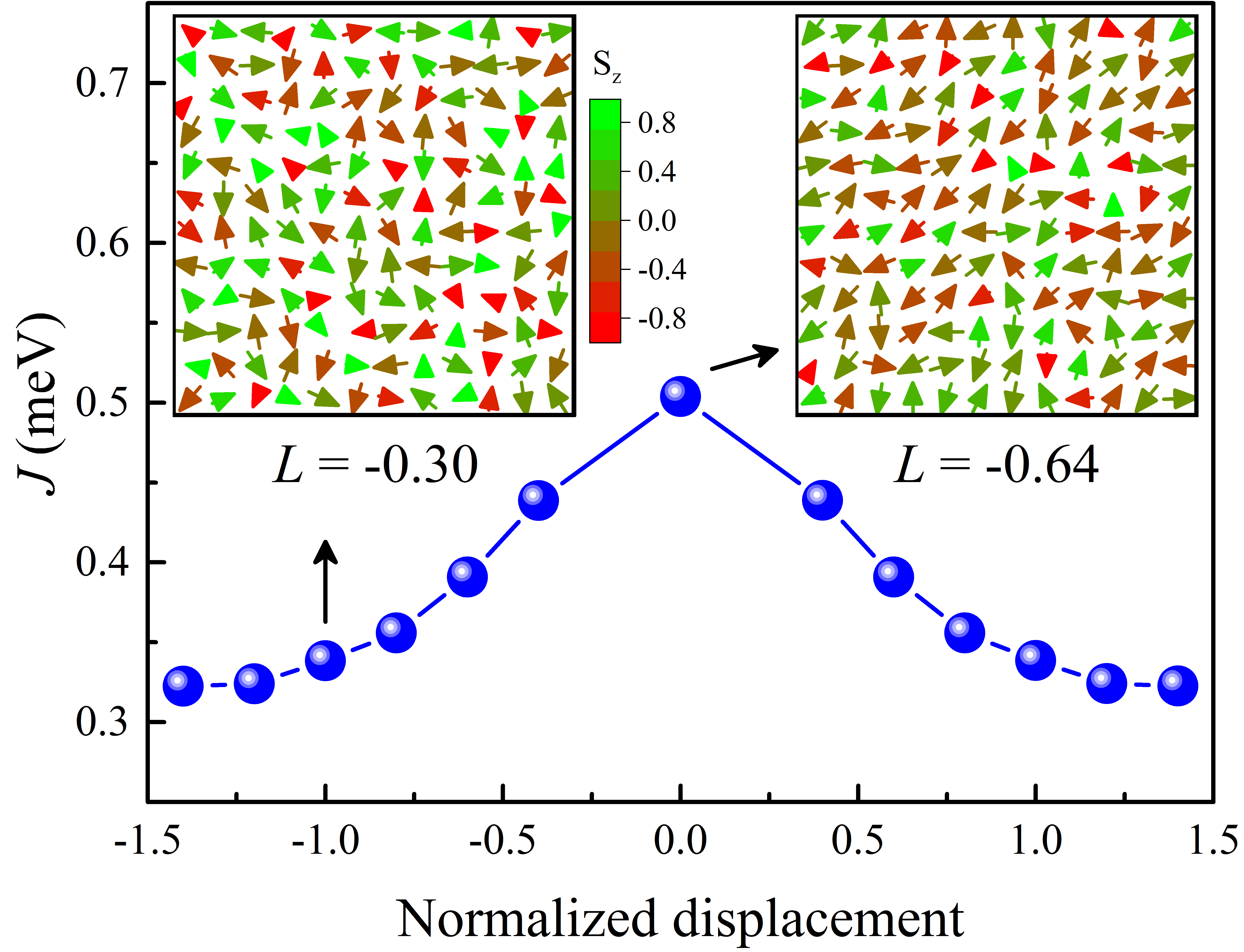}
\caption{Schematic of ME effect in the rhombohedral phase of GdWN$_3$. By suppressing the polar distortion, the magnetic coupling $J$ is enhanced by $\sim47\%$, which is $12$ times of that in BiFeO$_3$ ($\sim-3.8\%$ \cite{Xu2019a}). The insets are MC snapshots of spin configurations for $R3c$ (left) and $R\bar3c$ (right) states at $6$ K. The average antiferromagnetic correlation $L$ is defined as $<\mathbf{S}_i\cdot\mathbf{S}_j>$ (i.e. $L=-1$ for the ideal G-type antiferromagnet).}
\label{fig4}
\end{figure}

The superiority of those materials with same-ion-rooted ferroelectricity and magnetism is their inherent magnetoelectric coupling. To demonstrate this point, we calculate the exchange parameter $J$ as a function of normalized polar displacement in the rhombohedral phase, as shown in Fig.~\ref{fig4}. It should be noted that the material is in the paramagnetic state at ambient temperature, thus the magnetoelectricity is only discussed at low temperatures around $T_N$. As expected, the suppression of polar displacement will significantly enhance the magnetic coupling $J$ ($\sim47\%$ enhancement from $R3c$ to $R\bar3c$), which is more than one order of magnitude stronger than that in BiFeO$_3$ ($\sim-3.8\%$ for the same process \cite{Xu2019a}). Our MC snapshots (insets of Fig.~\ref{fig4}) also reveal a disorder-order tendency of spin texture tuned by the polar displacements, which further confirms the strong magnetoelectric coupling. In practice, this effect can occur in ferroelectric domain walls, where the local ferroelectric polarization is suppressed and thus the local magnetism is enhanced. Therefore, an external electric filed can switch the ferroelectric domains and tune the local magnetism at a certain temperature region.

This prominent magnetoelectricity is physically reasonable, since the polar displacements of Gd ion directly change the Gd-N-Gd bond length/angle, and thus reduce the orbital hybridization which is the source for magnetic exchange. In contrast, in BiFeO$_3$, the polar displacement is mainly contributed by Bi$^{3+}$, while the magnetic Fe$^{3+}$ ion is only passively involved. Thus the same-ion-rooted multiferroicity can provide stronger magnetoelectricity than those typical type-I multiferroics.

\begin{figure}
\centering
\includegraphics[width=0.48\textwidth]{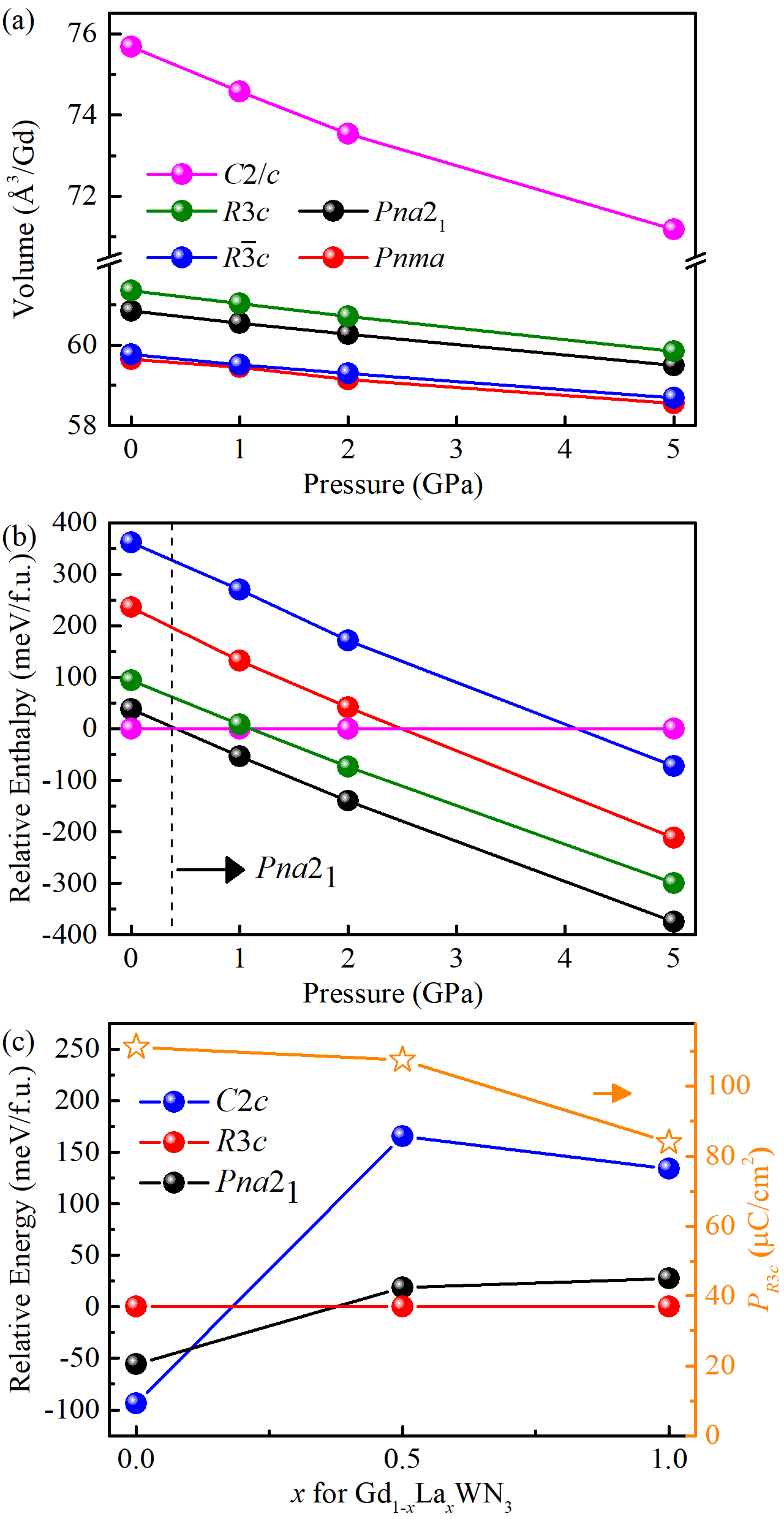}
\caption{(a) The volumes and (b) the enthalpies of different phases as a function of pressure. The enthalpy of $Pna2_1$ is set as the reference. (c) The polarization as a function of the in-plane strain. (d) The energies of various phases and the polarization vary with the Gd content $x$ in La$_{1-x}$Gd$_x$WN$_3$.}
\label{fig5}
\end{figure}

\subsection{Stabilize the perovskite phases}
Although $Pna2_1$ and $R3c$ GdWN$_3$ are interesting regard their multiferroic properties, they are energetically metastable. To stabilize them, the most convenient way is to use external pressure to induce a phase transition from the loose $C2/c$ structure to a compact perovskite structure. This method was once proved feasible in LaMoN$_3$ \cite{Gui2020}, and might be a general way to obtain metastable nitride perovskites.

By applying hydrostatic pressure to five most possible phases, their structures are further relaxed till the numerical deviation from the destination pressure smaller than $0.1$ GPa. Their volumes versus pressure curves are shown in Fig.~\ref{fig5}(a). It is clear that the original $C2/c$ cell is much larger ($24\%$ larger than the $Pna2_1$ one), and is softer upon pressure. Thus, the more compact perovskites might be more favorable under pressure.

Enthalpy is the criterion to determine the most stable structure under pressure at zero temperature. Thus the enthalpies versus pressure for various phases are plotted in Fig.~\ref{fig5}(b). A phase transition from $C2/c$ to $Pna2_1$ is expected at a small pressure $\sim0.5$ GPa, and since then the $Pna2_1$ phase always has the lowest enthalpy in the calculated range. Note that this is a first-order phase transition, so the $Pna2_1$ phase can keep metastable at the ambient condition after formation under pressure. The required pressure is small, which is easy to reach in experiment. More importantly, the pressure induce a $Pna2_1$ phase has not been stabilized in nitride perovskites before, while previously it was the $R3c$ phase stabilized by pressure in LaMoN$_3$ \cite{Gui2020}.

However, above pressure can only stabilize the $Pna2_1$ phase, while the more interesting $R3c$ phase remains unavailable. It is essential to recommend a proper experimental approach to obtain the $R3c$ phase. Since the sister member LaWN$_3$ owns the $R3c$ phase as the ground state \cite{Fang2017,Talley2021}, it is natural to expected that the partial substitution of A site Gd$^{3+}$ by La$^{3+}$ may be helpful. First, our calculation confirms the $R3c$ ground state of LaWN$_3$. Then, the simplest case, i.e., Gd$_{0.5}$La$_{0.5}$WN$_3$, is studied by comparing three most possible phases of Gd$_{0.5}$La$_{0.5}$WN$_3$, as shown in Fig.~\ref{fig5}(c). Luckily, the $R3c$ phase still has the lowest energy for this half-substituted case, while its polarization remains very large $107.6$ $\mu$C/cm$^2$. Moreover, significant enhancement of $J$ is also observed in this half-substituted case when the structure is switched from $R3c$ to $R\bar3c$, as shown in Fig.~S4 in SM \cite{supp}. Thus, following experiments are highly encouraged to study the Gd$_{1-x}$La$_{x}$WN$_3$ series.

\section{Conclusion}
In summary, an example of nitride perovskites, GdWN$_3$, has been studied with first-principles calculations to elucidate its intriguing multiferroicity with a gaint polarization and a large magnetic moment. Although its two multiferroic phases are meta-stable in energy, they can be obtained by pressure or ion-substitution. The nontrivial mechanism here is that the small size rare earth ion can contribute to both the ferroelectricity and magnetism, as a rare case of same-ion-rooted multiferroicity. Although only the Gd-case is studied herie, the underlying mechanisms should work for other rare earth cases, leading to more choices to pursuit high-performance multiferroics.

\begin{acknowledgments}
This work was supported by the National Natural Science Foundation of China (Grant No. 12104089). We thank the Big Data Center of Southeast University for providing the facility support on the numerical calculations.
\end{acknowledgments}

\bibliographystyle{apsrev4-2}
\bibliography{GdWN3}

\end{document}